# Bougainvillea derived porous carbons and their performance in magnetic field as supercapacitor electrodes

Nitish Yadav and S. Patnaik*
*School of Physical Sciences, Jawaharlal Nehru University, New Delhi (India)*
*E-mail: spatnaik@mail.jnu.ac.in

## Abstract

With the increase in demand of electrical energy storage devices such as batteries and supercapacitors, considerable effort is being put to increase the efficiency and applications of current technology while keeping it sustainable. Keeping this in mind we have persued the preparation and charatization of waste biomass derived carbon powders as supercapacitor/battery electrodes. Additionally, we have evaluated the performance of such carbons in the presence of an external magnetic field as we expect the graphene like structures to possess intrinsic magnetic nature. Here, we report the valorization of bougainvillea flower petals and detritus into graphenic carbon and explore a novel electrical double layer supercapacitor device that uses $Zn^{+2}$ ions for internal charge transport and is able to show increased performance upon application of an external magnetic field.

## 1 Introduction

Supercapacitors are energy storage devices with high power and energy density. They are generally categorized into two broad classes: EDLCs and pseudocapacitors. Pseudocapacitors store charge by fast-faradaic reactions at the surface or at a depth of a few atomic layers of the electrode active material.[1] While having higher values of specific capacitance compared to EDLCs and comparable power densities, they suffer from low cyclic stability due to faradaic reaction related surface-phase changes in the active material.[2] EDLCs, having physical charge-storage mechanism (double layer formation at electrode-electrolyte interface), are more attractive of the two types due to their simple construction and high stability, and due the possible use as complementary devices to the high energy dense, but less power dense, batteries.[3] The electrodes of EDLCs are generally made of highly porous carbons with a large surface area, having high durability, high colombic efficiency and long cylce life.

Porous carbons are versatile materials being applied in large number of applications such as water treatment[4], air filteration[5], seperation of chemicals[6], absorption of toxic compounds in alimantery canal[7], multisized pores and large porosity[8], reversible binding with metals and organic compound[9]. These properties make them suitable for use as electrodes in electrochemical energy storage devices such as electrical double layer capacitors (EDLCs) and lithium/potassium/sodium-ion batteries/capacitors.[10–17] Generally, porous carbons have two kinds of microstructures, the graphitic region, and the amorphous region.[18,19] The graphitic region is composed of small crystalline domains having multiple graphene-like layers stacked together in near-parallel arrangement with significant turbostratic disorder.[19] The microstructure of porous carbon particles can be classified according to the pore-size as well. There are mesopores (> 2 nm, < 50 nm) and micropores (< 2 nm) present, mostly in a hierarchical/fractally connected manner that are involved in ion-transport and ion-adsorption respectively.[18] This hierarchical/fractal connectivity enable proper transport of ions deep into the particles. Apart from these mesopores and micropores, another important charge-storage region is that between the graphene-like sheets of the quasi-crystalline/turbostratic/disordered regions of the porous carbons where ion storage happens via intercalation[20] (similar to that of lithium in graphite) at intermediate voltages (between 0.1 and 1 V). Such storage is important in devices such as batteries and metal-ion capacitors.[21,22] In both cases, i.e. adsorption on walls of micropores, and intercalation in graphitic domains, the nature of the interaction of the ion and the carbon structure plays an important role in defining the electrochemical stability (band-gap), leakage current, and enhancement in capacity.[22,23] While conventional research in energy storage devices aims to understand and improve the chemical, electrochemical, geometrical and electrical properties of the electrodes (carbon or otherwise), it ignores two important factors that can help enhance the charge-storage. The first of these less explored factors is the magnetic interactions between electrolyte and electrodes, and the second one is performance of the porous carbon electrodes in the presence of an external magnetic field ($B_{ext}$). The interaction between graphene, nano-graphene and other similar carbon structures with magnetic ions has been an interesting research topic which has been well studied by a number of groups and offers applications in diverse fields such as material separation and biomedical sensors.[24,25] The presence of $B_{ext}$ leads to magnetohydrodynamic (MHD) effect,[26] i.e. the change in the motion of ions in the electrolyte, inside the meso-/micropores and at the electrode-electrolyte interface in the presence of $B_{ext}$.[27] MHD effects caused by Lorentz force lead to uniform distribution of ions at the electrode-electrolyte interface, shorter diffusion paths and lower intefacial resistance.[26] The enhanced ionic diffusion in the porous electrode lead to higher rate capabilities along with better interfacial properties. In the current work for activation of the material during carbonization $ZnCl_2$ is chosen as the chemical activating agent. $ZnCl_2$ is not only a cost-effective, safe and recoverable chemical,[28–31] but is also easily able to dissolve cellulose present in the biomass (through breaking of interchain hydrogen bonds and formation of a zinc-cellulose complex[32,33]) leading to separation of molecular cellulose layers from the cellulose nanocrystallites which can then be easily converted to graphene-/graphite-like structures by thermal aromatization.[34–37] We study the graphenic porous carbons derived from waste biomass of the bougainvillae plant as electrode for storing $Zn^{+2}$ ion without and with the presence of an $B_{ext}$. In a three-electrode cell configuration, the application of the $B_{ext}$ is found to increase the performance of the electrodes. This enhancement can be attributed to the Lorentz force induced MHD effects on the cations ($Zn^{+2}$) and anions ($Cl^{-}$)as they move across the electrolyte into the porous regions of the carbon, and onto the electrode-electrolyte interface.

## 2 Experimental details

### 2.1 Materials

Bougainvillea petals and detritus were collected from Jawaharlal Nehru University campus, New Delhi, The activating agent $ZnCl_2$ (95% purity) was procured from Merck (India). Screen printed 3-electrode cells (glassy carbon working electrode with 3 mm dia., Ag/AgCl reference electrode, glassy carbon counter electrode) were procured from Aritech Chemazone, India.

**2.2 Preparation of activated carbon from Bougainvillea Petals**

Activated carbon was prepared from bougainvilleal flower petals by following the standard procedure described elsewhere and concisely shown in Fig. 1.[38] Bougainvillea flower petals were cut into small pieces using a pair of scissors. These pieces were then dried overnight at ~60 °C in a vacuum oven to obtain a dry and powdery mass. $ZnCl_2$ (chemical activating agent) was added to this dried powder in a ratio of 1 : x (w/w) (EuAC : $ZnCl_2$ ), where x: 0 and 2. Deionized water was added to form a slurry and stirred overnight for proper mixing. This slurry was dried again overnight at ~100 °C in the vacuum oven. The obtained mixture was put in alumina boats, which were then placed inside a tube furnace. The temperature of the furnace was increased from room temperature to 800 °C at a rate of 5 °C $min^{-1}$. Throughout this period, argon gas was purged in the furnace to provide an inert atmosphere for the simultaneous process of carbonization and chemical activation. Upon reaching 800 °C the furnace temperature was maintained at 800 °C for 2 hours. Thereafter, the furnace was allowed to cool down naturally while maintaining the flow of argon The resultant black mass was washed with dilute hydrochloric acid to remove residual $Zn^{+2}$ ions (in the case of x = 2) and then washed thoroughly with hot doubled distilled water by vacuum filtration until the pH of the filtrate reached the value of 7. The resulting mass was again dried in a vacuum oven at ~110 °C overnight to get the final (activated) carbon powder.

**2.3 Characterization of Activated Carbon**

Transmission electron microscope (TEM) images were recorded on a JEOL JEM-2100F instrument (Japan), set at a potential difference of 20 kV. The XRD measurements were performed on an x-ray diffractometer (Rigaku Miniflex600, Japan) using Cu-Kα radiation with a wavelength of 1.54 Å and 2Θ varying from 10 to 60 at a scan rate of 2  $min^{-1}$. Fourier-transform infrared (FTIR) spectroscopic studies were performed using a Perkin Elmer Frontier (UK) instrument in the wave number range from 450 to 4000 $cm^{-1}$. The spectra were recorded by averaging 32 scans per sample with an optical resolution of 1 $cm^{-1}$.

### 2.4 Electrochemical characterization of carbon materials in three-electrode assembly

A suspension of the active material (carbon) powder is prepared in ethanol (1 mg carbon for 1 ml ethanol) by ultrasonication for 30 min. 20 μL aliqouts of this suspension are then drop coated on the working electrode of the screen printed cell and kept for drying at room temperature for 30 min. This three-electrode cell is immersed inside a vial and 1 M $ZnCl_2$ (aq.) electrolyte filled in the vial such as to completely submerge the working, reference and counter electrodes of the cell. The three electrode cell is then used for electrochemical measurements in the absence/presence of magnetic field applied by a permanent magnet placed right next to the vial. A value of 1000 G was measured at the spot where the working electrode is placed using a gaussmeter. All measurements were performed at room temperature and ambient pressure. EIS, and CV studies have been carried out to characterize the carbon materials. EIS was performed from $10^5$ to $10^{-2}$ Hz at a signal level of 10 mV. EIS and CV were performed using the same electrochemical work-station (CHI660E, CH Instruments, USA).

## 3 Results and discussion

### 3.1 Structural and morphological analysis of carbon powders

Due to the difference in the biological composition of the bougainvillae petal and detritus material (detritus contains dried leaves along with dried petals) we expect difference in the micro/nano-structure of the resulting carbon powders. Fig. 2(a to h) show the TEM images for the carbon powders derived from bougainvillae petals and detritus at various magnifications. The presence of graphene nanodots and nanoplatelets is clearly visible (~ 5 - 250 nm long edges) in all images. While larger nanoplatelets (100 - 200 nm) are predominant in the BVPGC sample, the a-BVPGC sample has much smaller (5 - 20 nm size) nanodots and few nanoplatelets. Importantly, the TEM images of a-BVPGC show the presence of graphene-like nanoplatelets (100-200 nm size), on which are ’sprinkled’ a large number of smaller (approx. 5 nm - 10 nm), nanodots. This conclusion which matches well with our analysis of XRD data (Fig. 3a, discussed below) that the a-BVPGC sample has a narrow (002) peak (compared to BVPGC) indicative of graphene-like sheets arranged randomly over the larger platelets. BVDGC and a-BVDGC powders also have carbon nanorod like structures having width in the 5-10 nm range and lengths in the range 50-100 nm. It should be noted that there is a considerable number of carbon nanorods in BVDGC while only a trace amount of the same in a-BVDGC. Also present in the a-BVDGC sample are carbon nanoribbons of slightly larger width but similar length as the nanorods. The graphene nanoplatelets in BVDGC have a wide distrubtion of dimension (approx. 20 nm to a 250 nm), while those in a-BVDGC are smaller (approx. 20 nm - 100 nm). The smaller size of the nanostructures in a-BVPGC and a-BVDGC is due to the chemical etching of the carbon materials by activating agent $ZnCl_2$. Notably, the nanorods in a-BVDGC are formed by the carbonization of cellulose nanocrystallites.[34] In the case of BVDGC these are found to be connected to the edges of graphene-like sheets formed from the hemicellulose and lignin-rich[39,40] part of the raw material. The existence of the nanoribbons seen in a-BVDGC is possibly due to the complete/partial opening-up/dissolution of the cellulose nanocrystallites in the presence by $ZnCl_2$ as explained by Wustenberg et al.[41]. Cellulose nanocrystals consist of hexagonal arrangement of cellulose chains which can be divided into three regions, a central core region which is perfectly crystalline and two subcrystalline outer regions.[42] The width of cellulose nanocrystals generally lie in the range of 3 to 20 nm,[42] which is in confirmation with the width of the nanorods and nanoribbons observed in the TEM images here, underlining the origin of the latter for crystalline nanocellulose. Cellulose swelling and dissolution is a complex process and is completed by formation of concentric ring like rolls that open up into thin fragments.[41] $ZnCl_2$ has previously been reported to dissolve cellulose,[43] swell it, and increase its amorphousity.[28]

X-ray diffraction patterns of the porous carbon powders are shown in Fig. 3(a and b). A weak and broad peak, corresponding to the (002) plane is clearly seen at 25°. Another broad and barely visible hallow centered at 43.5°, corresponding to (101) family of planes, is present. The broad but weak nature of the peaks indicate towards the presence of few amorphous structures in the carbon powders, as well as the turbostratic arrangement of the graphene nanoplatelets and nanoribbons over one another. The (002) peak at 25° for a-BVPGC is noted to be narrower than that for BVPGC, a possible reason for which could be presence of numerous graphene-like structures in a random manner over each other and larger graphene-like platelets, as seen in Fig. 2(d). For the BVPGC and BVDGC samples, a number of sharper peaks are also visible corresponding to the (002) plane (18°, 23°, 26.6°, 28.4° and 29.4°), (101) plane (43.5°), and (100) plane (40.5°). The vast number of peaks for (002) plane are caused by the various nanostructured carbons present in the sample, each varying in its chemical functionalities, defects, and curvature and rotations of basal plane.[44] Similarly, the a-BVDGC sample shows clear crystalline peaks for 2Θ values of 20.8°, 26.6° and 28.5°, for the (002) planes and a weak peak at 52° corresponding to (004) reflection. The XRD plot for a-BVPGC is more indicative of a microstructure devoid of large-scale order, in agreement with our observations of the TEM images. The absence of sharp diffraction peaks corresponding to (101) and (100) planes indicates large scale graphitic regions are not present in a-BVPGC and a-BVDGC or that these planes either very small (in a-BVPGC), or curved, rotated and/or turbostratically disrupted (in a-BVDGC).

### 3.2 Infrared spectra of carbon powders

The presence of excess $sp^3$ (as well as $sp^2$) C-H bonds is also verified by comparing the infrared spectra of BVPGC, a-BVPGC, and BVDGC and a-BVDGC (Fig. 3c and 3d). BVDGC shows strong absorption bands, corresponding to various C-H vibrations, at ~1420 $cm^{-1}$, 880 $cm^{-1}$ and between 650 and 700 $cm^{-1}$, which are much weaker for a-BVDGC. Other than C-H bonds, signatures of other functionalities and adsorbed or interstitial moities present on the surface or edges are clearly visibile in the infrared spectra of both BVDGC and a-BVDGC. Important groups that are deduced from the infrared spectra are phenols, alcohols, ethers, nitriles, carbonyls, quinines, carboxyls, esters, carbonates, pyridines, pyrroles and amines and their corresponding bands are indicated.

### 3.3 Effect of external magnetic field on electrochemical measurements of bougainvillea petal and detritus derived carbons

To get an idea about the type of magnetic field strengths we should analyse our sample with one must note that the typical values of magnetic field strength in electric motors is of order of 1T (10,000 G).[45] and that of a hair dryer is of few tens of μT at distance of about 30 cm.[46,47] A laboratory scale helmholtz coil can produce magnetic fields with strength of ∼10 mT (100 G).[48] And the magnetic field near the Earth's poles is about 60 mT (0.6 G) and near the equator is ∼30 mT (300 G).[49] Thus, there are a number of ways in which the device can be exposed to magnetic fields of varying strengths. if we integrate supercapacitor devices just inside the casing of an electric motor, the supercapacitor will feel the effects of the magnetic fields of the order of 100 μT.[50] Magnetic field of such magnitude is expected to lead to MHD effects and/or variation in band structure of the electrode material. Additionaly, while acting as charge-storage devices, the supercapacitors can act as magnetic sensors if we record any deviations in their performance metrics, such as current vs voltage. This can be useful in scenarios where we experience regular fluctuations in magnetic field strength, such as a magnetic field medical/industrial device, or in satellites in space. Thus understanding the effect of $B_{ext}$ on the present supercapacitor electrodes, which can be used to fabricate solid state and flexible devices is important. Flexible, easily shaped supercapacitor devices (as reported in a number of publications[51,52]) can be adjusted in a large number of geometries, from distorted volumes to planar, flexible surfaces such as textiles. We have tested the prepared carbon electrodes in three electrode screen-printed cells. CV plots have been recorded at a scan rate of 20 mV $s^{-1}$ in the voltage range of 0 - 1 V.

The comparative CV plots for the four carbon electrodes are shown in Fig.4 (a to d). For the bougainvillae petal derived carbon samples (BVPGC and a-BVPGC) a slight increase in current density in the presence of $B_{ext}$ is observed. The increase in current density is explained by the Lorentz force on the mobile ions which imparts a lateral velocity component to the ions adsorbed on the electrode-electrolyte interface. The shape of the CV plots for BVPGC and a-BVPGC is also more rectangular in the presence of $B_{ext}$ due to the lowered resistance to double-layer formation. For the case of the detritus derived samples, the $B_{ext}$ has no effect at all on the performance of the electrodes, indicating that the microstructure of the carbon too plays an important role in determining the effect of $B_{ext}$ on device/electrode performance. Indeed, as seen in the TEM images, the detritus derived carbons have large number of graphene nanoribbons or platelets. This is also the case for BVPGC, which is composed of graphene nanoplatelets. In the CV plots, a-BVPGC electode has a current response almost 4-5 times higher than that for the other three samples. The large current response for a-BVPGC is due to the greater surface available for the ions to form an electrode-electrolyte interface. The anamolous behaviour of BVPGC, which has a low current response like BVDGC and a-BVDGC, but one that improves in the presence of $B_{ext}$, could be due to the well-formed graphene nanoplatelets, and the low migration energy of $Zn^{+2}$ ions over these nanoplatelets[53] coupled with their low number density. The low migration energy for $Zn^{+2}$ and the large surface of a nanoplatelet allows for large scale diffusion of $Zn^{+2}$ across a single nanoplatelet, in the presence of $B_{ext}$, before encountering a boundary/edge, ensuring formation of a better double-layer compared to the case without $B_{ext}$. For the BVDGC and a-BVDGC samples $B_{ext}$ is unable to impart significant improvement due to more amorphous content of BVDGC (as revealed in XRD analysis) and the insufficient size of the nanorods and nanoribbons in a-BVDGC, for lateral ion motion due to $B_{ext}$ to be significant.

The conclusions from the analysis of the CV plots can also be confirmed by the analysis of the a.c. impedance data. For this, the frequency dependent real and imaginary impedance have been plotted (Fig. 5a to d) and fitted (Fig. 6a to d) with the equivalent circuit shown in Fig. 6(e) and the fit parameters listed in Table 1. The equivalent circuit consists of $R_b$ and $R_{ct}$, representing the bulk resistance and resistance associated with charge-transfer from ions at the electrode-electrolyte interface, respectively. A constant phase element $CPE_{ads}$ is used in the equivalent circuit to represent an imperfect capacitive element modeled by the equation:[54]

$$CPE = C_\alpha/(j\omega)^\alpha \qquad (1)$$

The $R_{ct}$ is placed in series with $W_o$, a open Warburg element, representing the diffusion controlled ion motion through the highly inhomogenous microstructure between the micro/nano-sized spaces between the graphene nanostructures and represented by the equation:

$$W_o = (A/\sqrt{(j\omega\tau)})*\coth\sqrt{(j\omega\tau)} \qquad (2)$$

The value of α (for CPE) varies between 0 and 1, and $C_\alpha$ represents an ideal capacitor for $\alpha = 1$. For $\alpha < 1$, *CPE* is representative of a non-homogenous system with a wide distribution of time constants and has a frequency-independent phase angle different from 90°.[54,55] For the case of $W_o$, A is the Warburg coefficient, $\tau = L^2/D$ represents the characteristic time constant, $L$ is the typical pore length and $D$ is the diffusion constant for ion in pore.[56] Few important conclusions can be drawn from the values of the parameters listed in Table 1. The value of $R_b$ shows almost constant value for the four samples without and with $B_{ext}$. It increases slightly from 78 Ω $cm^2$ to 109 Ω $cm^2$ for the case of BVPGC (which can be considered to be almost same value as the scale of the measurement value in this case is almost 100 times that of the difference between the two values), remains the same for a-BVPGC (191 Ω $cm^2$), decreases slightly for BVDGC (from 340 Ω $cm^2$ to 263 Ω $cm^2$) and only slightly for a-BVDGC (from 257 Ω $cm^2$ to 251 Ω $cm^2$). Thus, all these changes can be considered insignificant as already explained for the case of BVPGC. The value of $R_{ct}$, which is an important parameter characterizing the electrode-electrolyte interface, varies significantly for the case of BVPGC and BVDGC for the cases without and with $B_{ext}$. However, no significant change is observed for the cases where the carbons have been activated. These differences in response to $B_{ext}$ can be understood by comparing the TEM images of the four samples. While the samples that show significant changes in $R_{ct}$ in the presence of $B_{ext}$ (i.e. BVPGC and BVDGC) have larger nanoplatelets, the a-BVPGC and a-BVDGC samples consists of smaller nanoribbons and nanodots. $R_{ct}$ is the resistance attributed to an electron's movement from a molecule/ion in the electrolytic to the carbon electrode. The larger value of $R_{ct}$ in the presence of $B_{ext}$ highlights the MHD effects on ion motion near the graphenic planes. The transverse component of ion velocity imparted by $B_{ext}$ makes it difficult for the ions to settle on the planes, and hence lowers the chance of charge-transfer. The MHD effect is not so effective for the case of the activated carbon samples as the smaller consitutent nanoribbons and nanodots have smaller surfaces which are futher aligned in all directions with almost equal probability. The coefficient of the $W_o$ element, $A$ and the characteristic time constant $\tau$ similarly show a drastic increase in presence of $B_{ext}$ for the unactivated carbon samples (BVPGC and BVDGC), while only nominal changes are observed for the activated samples. These changes too can be attributed to the large MHD effects in the carbon samples containing large sheet-like structures. The effects of $B_{ext}$ on capacitive double layer formation at the electrode-electrolyte interface are reflected in the $CPE$ element. The exponent $\alpha$ increases significantly from 0.40 to 0.72 when $B_{ext}$ is applied for the case of BVPGC, while the increase finite but much lower for the case of BVDGC sample (from 0.70 to 0.74). The value of $C_\alpha$ on the other hand increases by an order of 10 for BVPGC when $B_{ext}$ is applied. For BVDGC, we again see a slighter but finite increase. The increase in values of $C_\alpha$ and $\alpha$ and for the unactivated samples in presence of $B_{ext}$ indicates better double layer formation and better penetration of the porous structure by the ions from the electrolyte. This is again attributed to the addition transverse component of velocity the ions moving in the pores gain due to the Lorentz force from $B_{ext}$, enabling better diffusion over hinderances and around corners in the porous network. As expected for the activated samples, there is no change in the value of either $C_\alpha$ or $\alpha$.

## 4 Conclusions

We have used waste biomass from bougainvillea shrub to prepare graphinic carbon powders which have also been chemically activated with $ZnCl_2$. The morphological and spectroscopic analysis of the prepared carbons reveal presence of graphene like nanostructures with functional groups attached. While the unactivated carbons have large graphene nanoplatelets, their activated counterparts exhibit smaller nanoribbons and nanodots, thus enabling us a contrast in our measurements of electrochemical properties. The cyclic voltammetric and a.c. impedance measurements carried out without and with the presence of an external magnetic field and with $Zn^{+2}$ containing electrolyte are influenced by the nanostructure of the carbon powders, with the unactivated samples containing large graphinic nanoplatelets being affected the most due to magnetohydrodynamic effects. The reported carbon powders can act as efficient energy storage materials not only for $Zn^{+2}$ based electrolytes but other sustainable technologies and their performance can be enhanced by appropriately incorporating them in regions of high magnetic fields found in daily household and industrial equipments as well as space based applications, where they can play the dual role of energy storage materials and magnetic field sensors.

## Acknowledgements

NY thanks Science and Engineering Research Board, New Delhi for funding through National Post-doctoral Fellowship award (file no. PDF/2019/001636).

## Conflict of Interests

There are no conflicts of interest to declare.

## References

1. Gogotsi, Y. & Penner, R. M. Energy Storage in Nanomaterials – Capacitive, Pseudocapacitive, or Battery-like? *ACS Nano* **12**, 2081–2083 (2018).
2. Teng, Y. *et al.* Bean dregs-based activated carbon/copper ion supercapacitors. *Electrochimica Acta* **194**, 394–404 (2016).
3. Kim, S. J. *et al.* Charging-Discharging Behavior and Performance of AGM Lead Acid Battery/EDLC Module for x-HEV. *Korean J. Mater. Res.* **31**, 84–91 (2021).

4. Kumar N, S., Grekov, D., Pré, P. & Alappat, B. J. Microwave mode of heating in the preparation of porous carbon materials for adsorption and energy storage applications – An overview. *Renew. Sustain. Energy Rev.* **124**, 109743 (2020).
5. Li, Z., Xu, J., Sun, D., Lin, T. & Huang, F. Nanoporous Carbon Foam for Water and Air Purification. *ACS Appl. Nano Mater.* **3**, 1564–1570 (2020).
6. Siegelman, R. L., Kim, E. J. & Long, J. R. Porous materials for carbon dioxide separations. *Nat. Mater.* **20**, 1060–1072 (2021).
7. Day, G. S., Drake, H. F., Zhou, H.-C. & Ryder, M. R. Evolution of porous materials from ancient remedies to modern frameworks. *Commun. Chem.* **4**, 1–4 (2021).
8. Borenstein, A. *et al.* Carbon-based composite materials for supercapacitor electrodes: a review. *J. Mater. Chem. A* **5**, 12653–12672 (2017).
9. Wilczak, A. & Keinath, T. M. Kinetics of sorption and desorption of copper(II) and lead (II) on activated carbon. *Water Environ. Res.* **65**, 238–244 (1993).
10. Delaporte, N. *et al.* Facile formulation and fabrication of the cathode using a self-lithiated carbon for all-solid-state batteries. *Sci. Rep.* **10**, 11813 (2020).
11. Etacheri, V., Wang, C., O'Connell, M. J., Chan, C. K. & Pol, V. G. Porous carbon sphere anodes for enhanced lithium-ion storage. *J. Mater. Chem. A* **3**, 9861–9868 (2015).
12. Niu, J. *et al.* Porous carbon electrodes with battery-capacitive storage features for high performance Li-ion capacitors. *Energy Storage Mater.* **12**, 145–152 (2018).
13. Cui, Y. *et al.* All-carbon lithium capacitor based on salt crystal-templated, N-doped porous carbon electrodes with superior energy storage. *J. Mater. Chem. A* **6**, 18276–18285 (2018).
14. Hu, X. *et al.* Hierarchical porous carbon nanofibers for compatible anode and cathode of potassium-ion hybrid capacitor. *Energy Environ. Sci.* **13**, 2431–2440 (2020).
15. Zhou, C. *et al.* Three-dimensional porous carbon doped with N, O and P heteroatoms as high-performance anode materials for sodium ion batteries. *Chem. Eng. J.* **380**, 122457 (2020).
16. Liu, X. *et al.* High-energy sodium-ion capacitor assembled by hierarchical porous carbon electrodes derived from Enteromorpha. *J. Mater. Sci.* **53**, 6763–6773 (2018).
17. Kim, M.-H., Kim, K.-B., Park, S.-M. & Roh, K. C. Hierarchically structured activated carbon for ultracapacitors. *Sci. Rep.* **6**, 21182 (2016).
18. Guo, F. *et al.* Synthesis of biomass carbon electrode materials by bimetallic activation for the application in supercapacitors. *J. Electroanal. Chem.* **844**, 105–115 (2019).
19. Stevens, D. A. & Dahn, J. R. The mechanisms of lithium and sodium insertion in carbon materials. *J. Electrochem. Soc.* **148**, A803 (2001).
20. Anji Reddy, M., Helen, M., Groß, A., Fichtner, M. & Euchner, H. Insight into sodium insertion and the storage mechanism in hard carbon. *ACS Energy Lett.* **3**, 2851–2857 (2018).
21. M. Rathnayake, R. M. N., T. Duignan, T., J. Searles, D. & S. Zhao, X. Exploring the effect of interlayer distance of expanded graphite for sodium ion storage using first principles calculations. *Phys. Chem. Chem. Phys.* **23**, 3063–3070 (2021).
22. Ma, J. *et al.* Improvement of alkali metal ion batteries via interlayer engineering of anodes: from graphite to graphene. *Nanoscale* **13**, 12521–12533 (2021).
23. Meng, J. *et al.* Advances in Structure and Property Optimizations of Battery Electrode Materials. *Joule* **1**, 522–547 (2017).
24. Hill, E. W., Vijayaragahvan, A. & Novoselov, K. Graphene Sensors. *IEEE Sens. J.* **11**, 3161–3170 (2011).
25. Yan, M. *et al.* Amino acid-modified graphene oxide magnetic nanocomposite for the magnetic separation of proteins. *RSC Adv.* **7**, 30109–30117 (2017).

26. Wang, G. *et al.* Facile synthesis of highly photoactive α-Fe2O3-based films for water oxidation. *Nano Lett.* **11**, 3503–3509 (2011).
27. Aogaki, R., Fueki, K. & Mukaibo, T. Application of Magnetohydrodynamic Effect to the Analysis of Electrochemical Reactions 1.　MHD Flow of an Electrolyte Solution in an Electrode－Cell with a short Rectangular Channel. 電気化学および工業物理化学 **43**, 504–508 (1975).
28. Burger, D. *et al.* Partial amorphization of cellulose through zinc chloride treatment: A facile and sustainable pathway to functional cellulose nanofibers with flame-retardant and catalytic properties. *ACS Sustain. Chem. Eng.* **8**, 13576–13582 (2020).
29. Solhy, A., Clark, J. H., Tahir, R., Sebti, S. & Larzek, M. Transesterifications catalysed by solid, reusable apatite–zinc chloride catalysts. *Green Chem.* **8**, 871–874 (2006).
30. PubChem. Zinc chloride. https://pubchem.ncbi.nlm.nih.gov/compound/5727.
31. Bi, Z., Lai, B., Zhao, Y. & Yan, L. Fast disassembly of lignocellulosic biomass to lignin and sugars by molten salt hydrate at low temperature for overall biorefinery. *ACS Omega* **3**, 2984–2993 (2018).
32. Leipner, H., Fischer, S., Brendler, E. & Voigt, W. Structural changes of cellulose dissolved in molten salt hydrates. *Macromol. Chem. Phys.* **201**, 2041–2049 (2000).
33. Cao, N.-J., Xu, Q., Chen, C.-S., Gong, C. S. & Chen, L. F. Cellulose hydrolysis using zinc chloride as a solvent and catalyst. *Appl. Biochem. Biotechnol.* **45**, 521–530 (1994).
34. Klemm, D., Philpp, B., Heinze, T., Heinze, U. & Wagenknecht, W. *Comprehensive cellulose chemistry. Volume 1: Fundamentals and analytical methods.* (Wiley-VCH Verlag GmbH, 1998).
35. Klemm, D., Philpp, B., Heinze, T., Hewinze, U. & Wagenknecht, W. *Comprehensive cellulose chemistry. Volume 2: Functionalization of cellulose.* (Wiley-VCH Verlag GmbH, 1998).
36. ROSS, S. CARBON AND GRAPHITE YARNS FROM VISCOSE RAYON. *Am. Dyest. Report.* **58**, 26 (1969).
37. Vohler, O., Reiser, P.-L., Martina, R. & Overhoff, D. New forms of carbon. *Angew. Chem. Int. Ed. Engl.* **9**, 414–425 (1970).
38. Hu, Z., Srinivasan, M. P. & Ni, Y. Preparation of mesoporous high-surface-area activated carbon. *Adv. Mater.* **12**, 62–65 (2000).
39. Salem, M. Z. M. *et al.* Antifungal Activities of Wood and Non-Wood Kraft Handsheets Treated with Melia azedarach Extract Using SEM and HPLC Analyses. *Polymers* **13**, 2012 (2021).
40. El-Chaghaby, G., Rashad, S. & Abd-ElKader, S. Dried Leaves of Bougainvillea glabra Plant for the Removal o Lead Ions from Aqueous Solution by Adsorption. *Egypt. J. Bot.* **60**, 707–718 (2020).
41. Wüstenberg, T. *Cellulose and cellulose derivatives in the food industry: fundamentals and applications*. (John Wiley & Sons, 2014).
42. Festucci-Buselli, R. A., Otoni, W. C. & Joshi, C. P. Structure, organization, and functions of cellulose synthase complexes in higher plants. *Braz. J. Plant Physiol.* **19**, 1–13 (2007).
43. El Oudiani, A., Chaabouni, Y., Msahli, S. & Sakli, F. Crystal transition from cellulose I to cellulose II in NaOH treated Agave americana L. fibre. *Carbohydr. Polym.* **86**, 1221–1229 (2011).
44. Li, Z. Q., Lu, C. J., Xia, Z. P., Zhou, Y. & Luo, Z. X-ray diffraction patterns of graphite and turbostratic carbon. *Carbon* **45**, 1686–1695 (2007).
45. Kovac, D. & Kovacova, I. The magnetic fields of electric motors and their EMC. *Adv. Electr. Electron. Eng.* **7**, 183–186 (2011).

46. Kaune, W. t. *et al.* Magnetic fields produced by hand held hair dryers, stereo headsets, home sewing machines, and electric clocks. *Bioelectromagnetics* **23**, 14–25 (2002).
47. Portal, E.-E. E. Concerns over electromagnetic fields below three-phase high voltage overhead lines - EEP. *EEP - Electrical Engineering Portal* https://electrical-engineering-portal.com/electromagnetic-fields-hv-overhead-lines (2018).
48. Fite, M. C., Rao, J.-Y. & Imae, T. Effect of external magnetic field on hybrid supercapacitors of nitrogen-doped graphene with magnetic metal oxides. *Bull. Chem. Soc. Jpn.* **93**, 1139–1149 (2020).
49. Earth's magnetic field. https://web.ua.es/docivis/magnet/earths_magnetic_field2.html.
50. Fireteanu, V., Taras, P. & Stamate, M. Finite element analysis and diagnosis of the one broken bar fault in the squirrel cage induction motors. in *2012 13th International Conference on Optimization of Electrical and Electronic Equipment (OPTIM)* 411–416 (IEEE, 2012).
51. Shao, Y. *et al.* Graphene-based materials for flexible supercapacitors. *Chem. Soc. Rev.* **44**, 3639–3665 (2015).
52. Peng, X., Peng, L., Wu, C. & Xie, Y. Two dimensional nanomaterials for flexible supercapacitors. *Chem. Soc. Rev.* **43**, 3303–3323 (2014).
53. Olsson, E., Hussain, T., Karton, A. & Cai, Q. The adsorption and migration behavior of divalent metals (Mg, Ca, and Zn) on pristine and defective graphene. *Carbon* **163**, 276–287 (2020).
54. Holm, S., Holm, T. & Martinsen, Ø. G. Simple circuit equivalents for the constant phase element. *PloS One* **16**, e0248786 (2021).
55. EIS Spectrum Analyser Help. Equivalent Circuit Elements and Parameters. http://www.abc.chemistry.bsu.by/vi/analyser/parameters.html.
56. Muralidharan, V. S. Warburg impedance-basics revisited. *Anti-Corros. Methods Mater.* (1997).

## Tables

**Table 1**: Parameters corresponding to the Equivalent Randle's circuit fit to the electrochemical (a.c.) impedance spectroscopy analysis data for the four carbon samples

| Electrode | *B* | $R_b$ ($\Omega$ cm$^2$) | $R_{ct}$ ($\Omega$ cm$^2$) | $W_o$ ($A/\Omega$ cm$^2$, $\tau$/s) | *CPE* ($C_\alpha/\Omega$ cm$^2$s$^{-\alpha}$, $\alpha$) |
|---|---|---|---|---|---|
| BVPGC | no | 78 | 38 | 151, $4\times10^{-5}$ | $1.3\times10^{4}$, 0.40 |
| BVPGC | yes | 109 | 2020 | 42500, 6 | $1.4\times10^{5}$, 0.72 |
| a-BVPGC | no | 191 | 564 | 288, $3\times10^{-3}$ | $3.3\times10^{3}$, 0.48 |
| a-BVPGC | yes | 191 | 509 | 86, $1\times10^{-3}$ | $3.4\times10^{3}$, 0.48 |
| BVDGC | no | 340 | 110 | 776, $1\times10^{-3}$ | $2.1\times10^{4}$, 0.70 |
| BVDGC | yes | 263 | 41100 | $1\times10^{5}$, $5\times10^{-3}$ | $2.2\times10^{4}$, 0.74 |
| a-BVDGC | no | 257 | 91 | 5, $4\times10^{-6}$ | $5.1\times10^{4}$, 0.64 |
| a-BVDGC | yes | 251 | 88 | 12, $9\times10^{-6}$ | $5.7\times10^{4}$, 0.65 |

## Figures

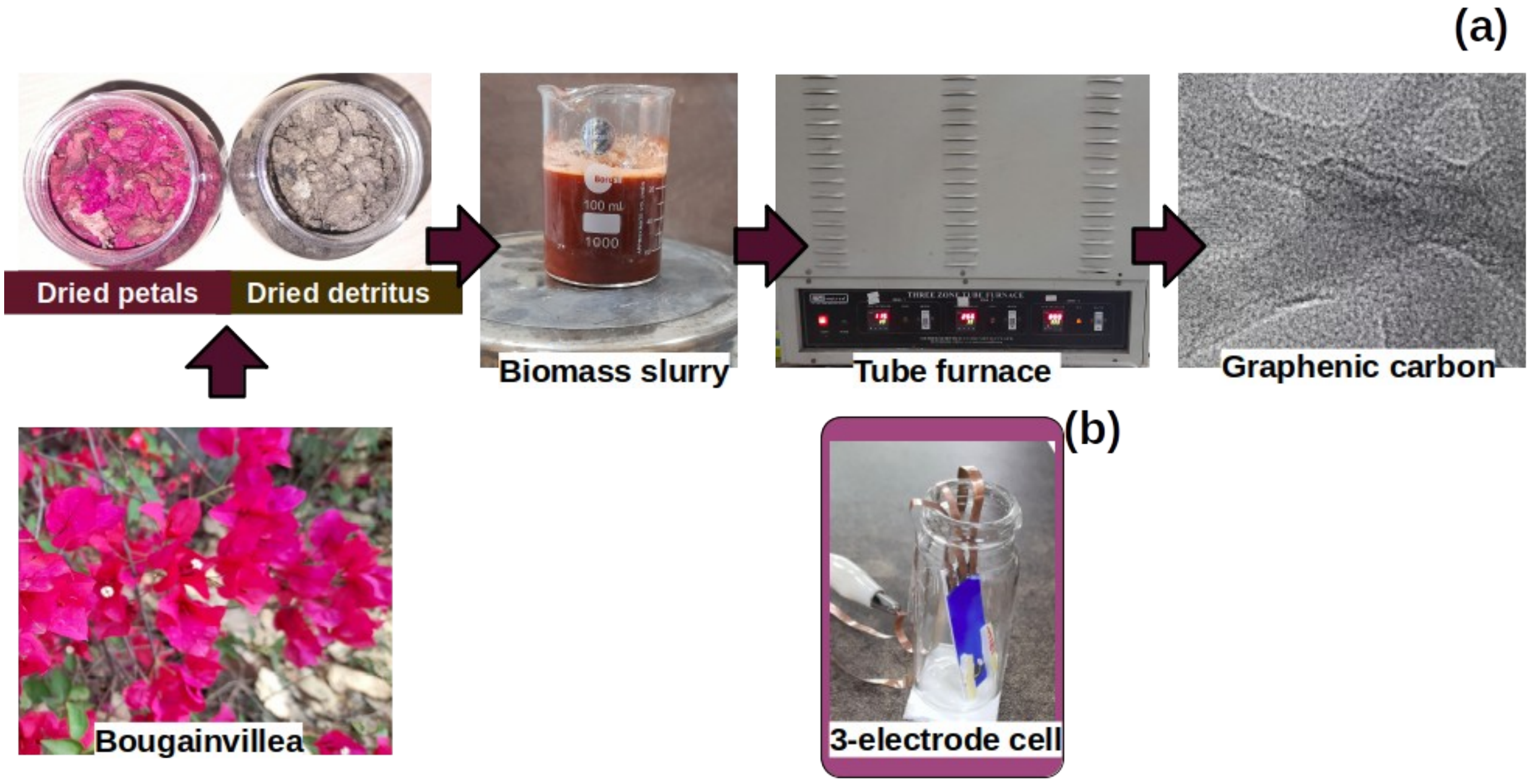

**Fig. 1**: (a) Schematic illustrating the synthesis procedure for bougainvillea petal and detritus derived graphenic carbon, and (b) screen-printed 3-electrode assembly for electrochemical measurements

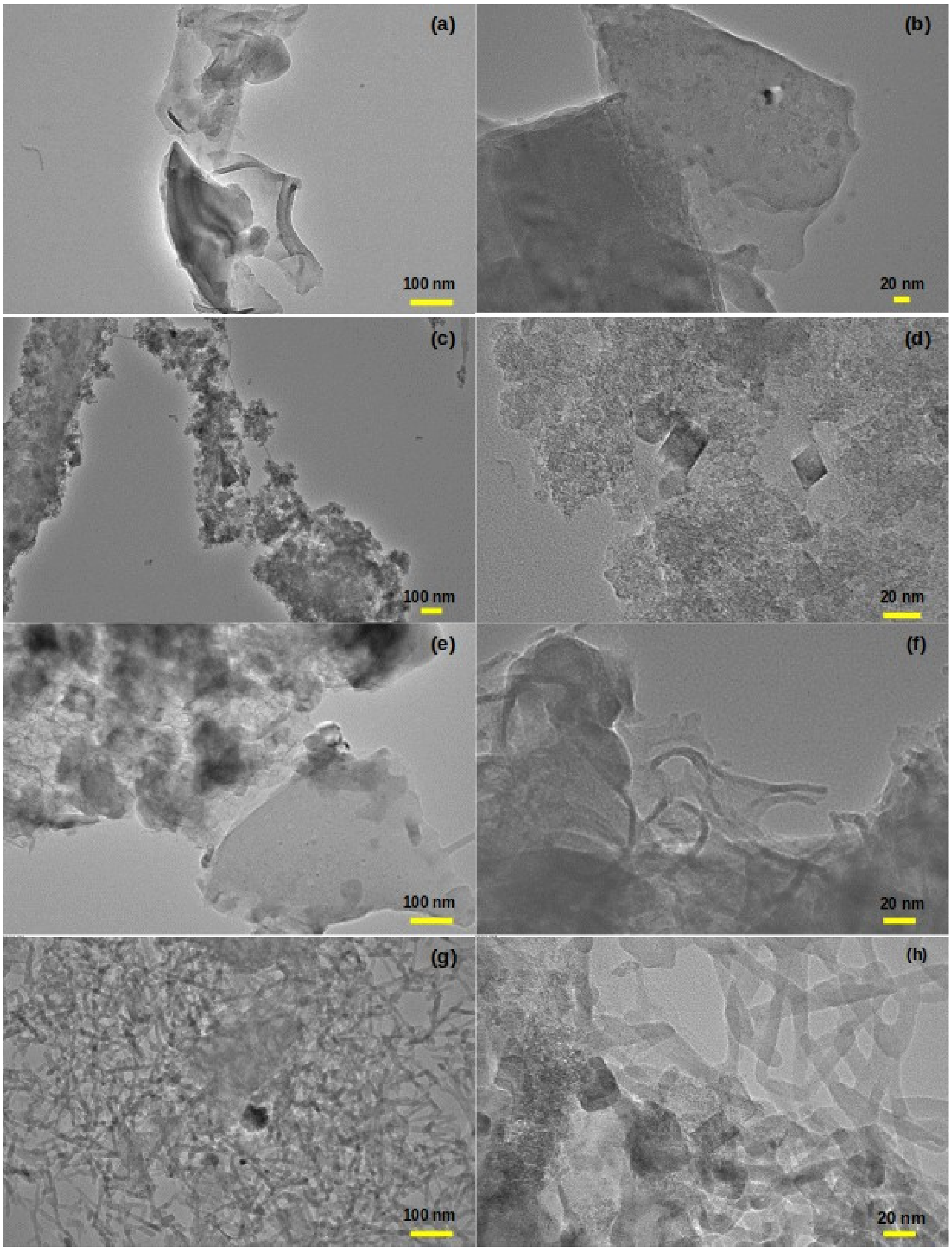

**Fig. 2**: TEM images of (a, b) BVPGC, (c, d) a-BVPGC, (e, f) BVDGC and (g, h) a-BVDGC.

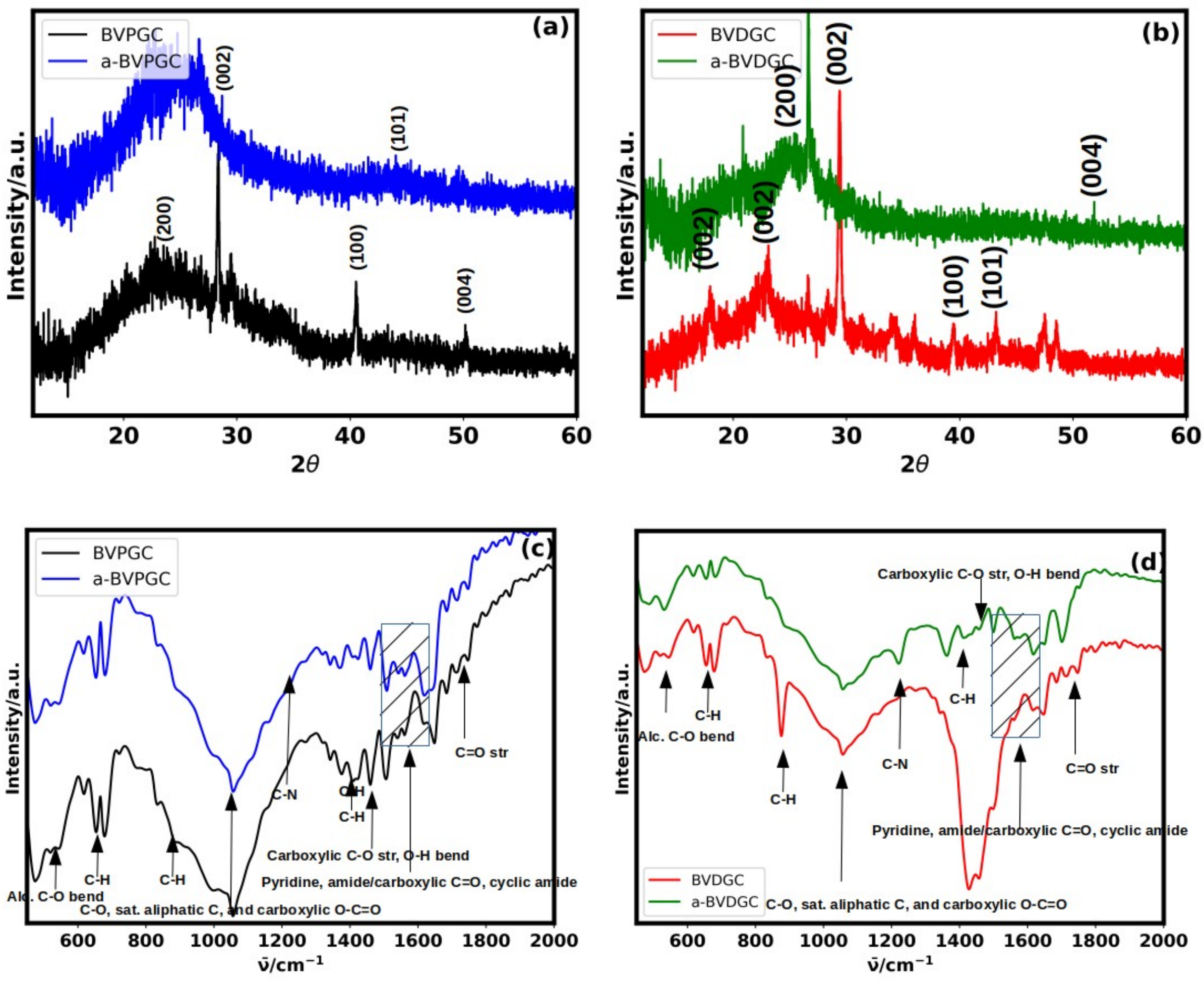

**Fig. 3**: X-ray diffractograms of BVPGC and a-BVPGC (a) and BVDGC and a-BVDGC (b). Infrared spectrograms of BVPGC and a-BVPGC (c) and BVDGC and a-BVDGC (d).

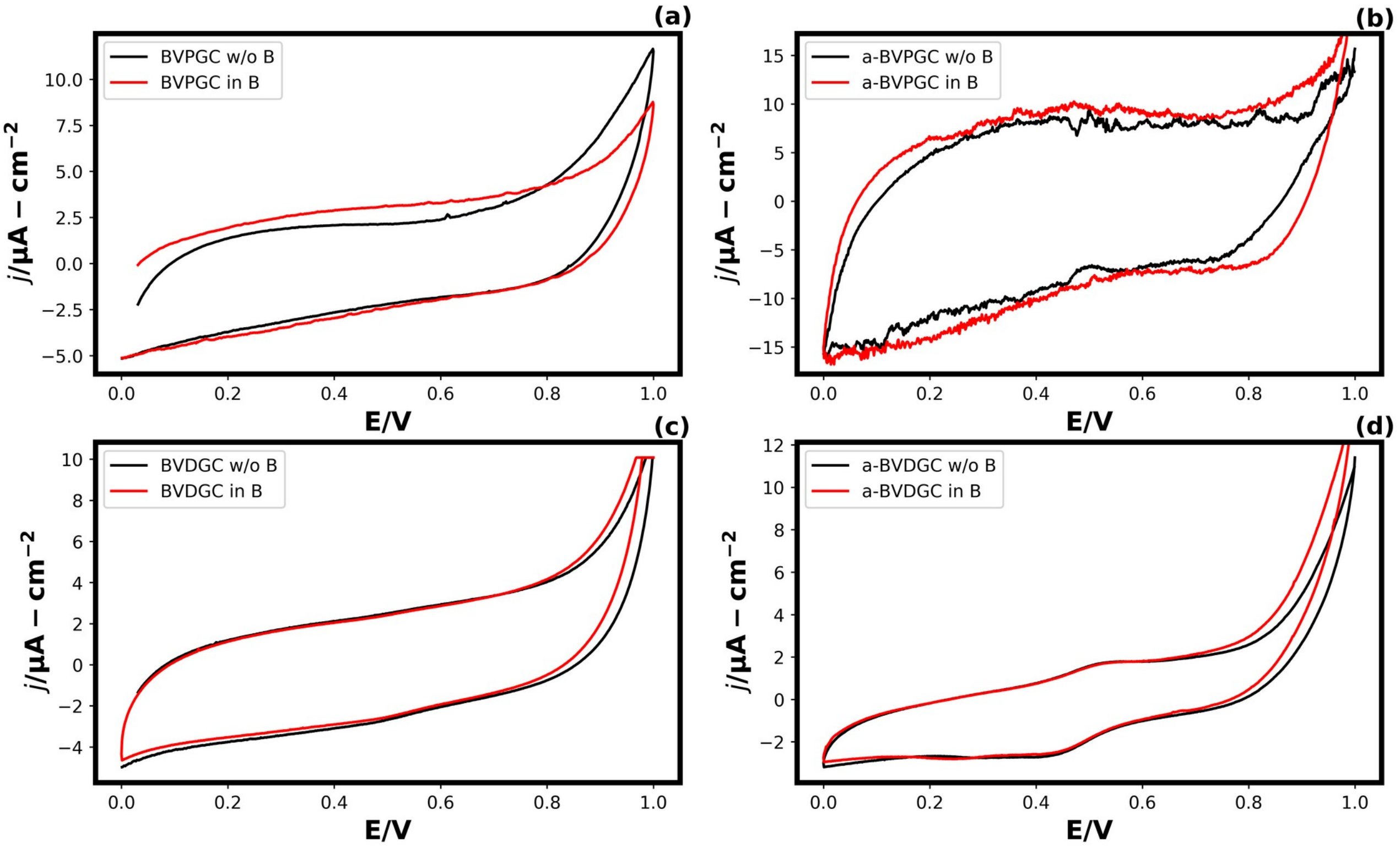

**Fig. 4**: (a-d) Cyclic voltammetric (CV) plots for the three electrodes cells fabricated using bougainvillea derived carbons.

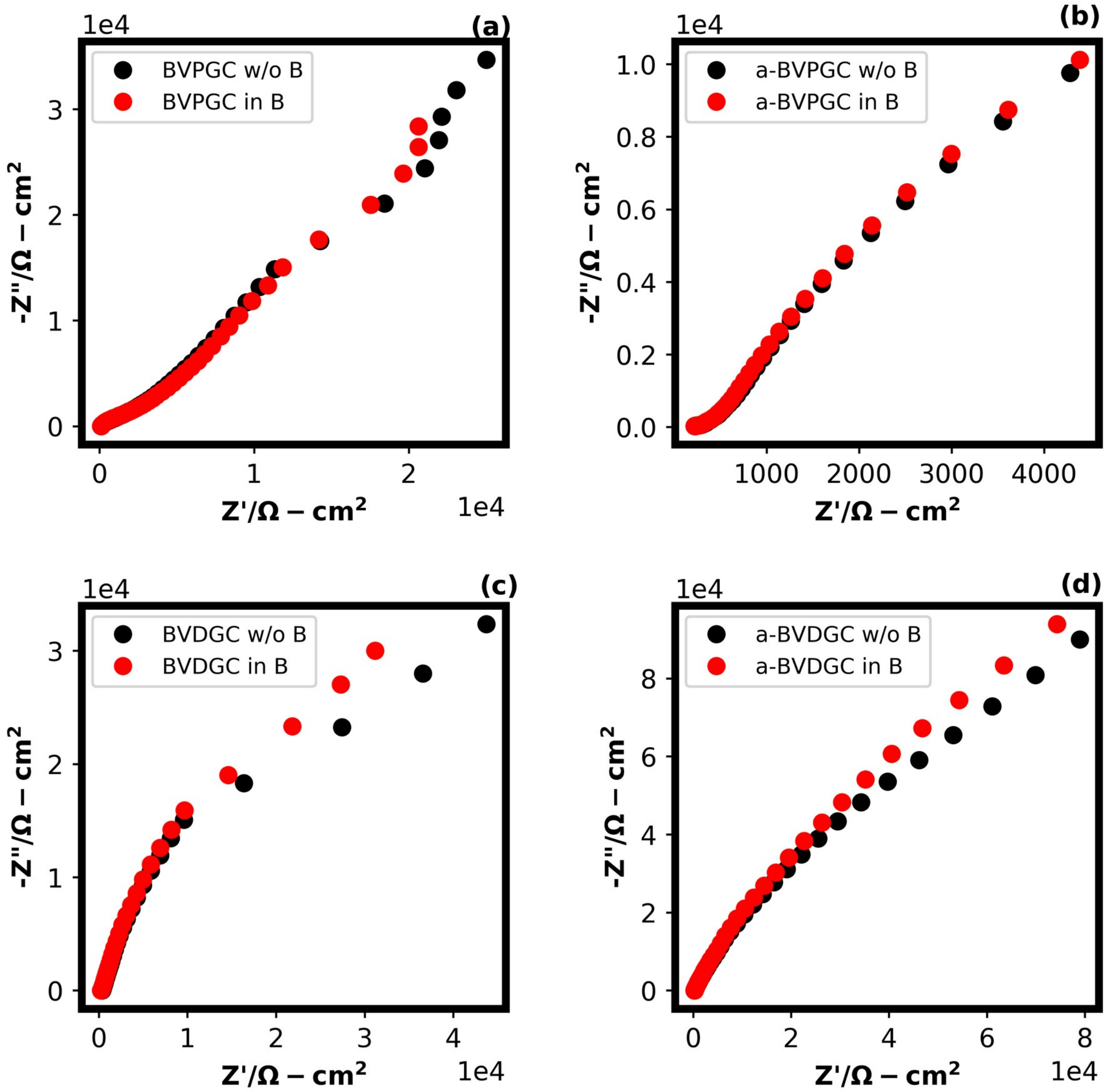

**Fig. 5**: (a-d) Nyquist plots of the three electrodes cells fabricated using bougainvillea derived porous carbons.

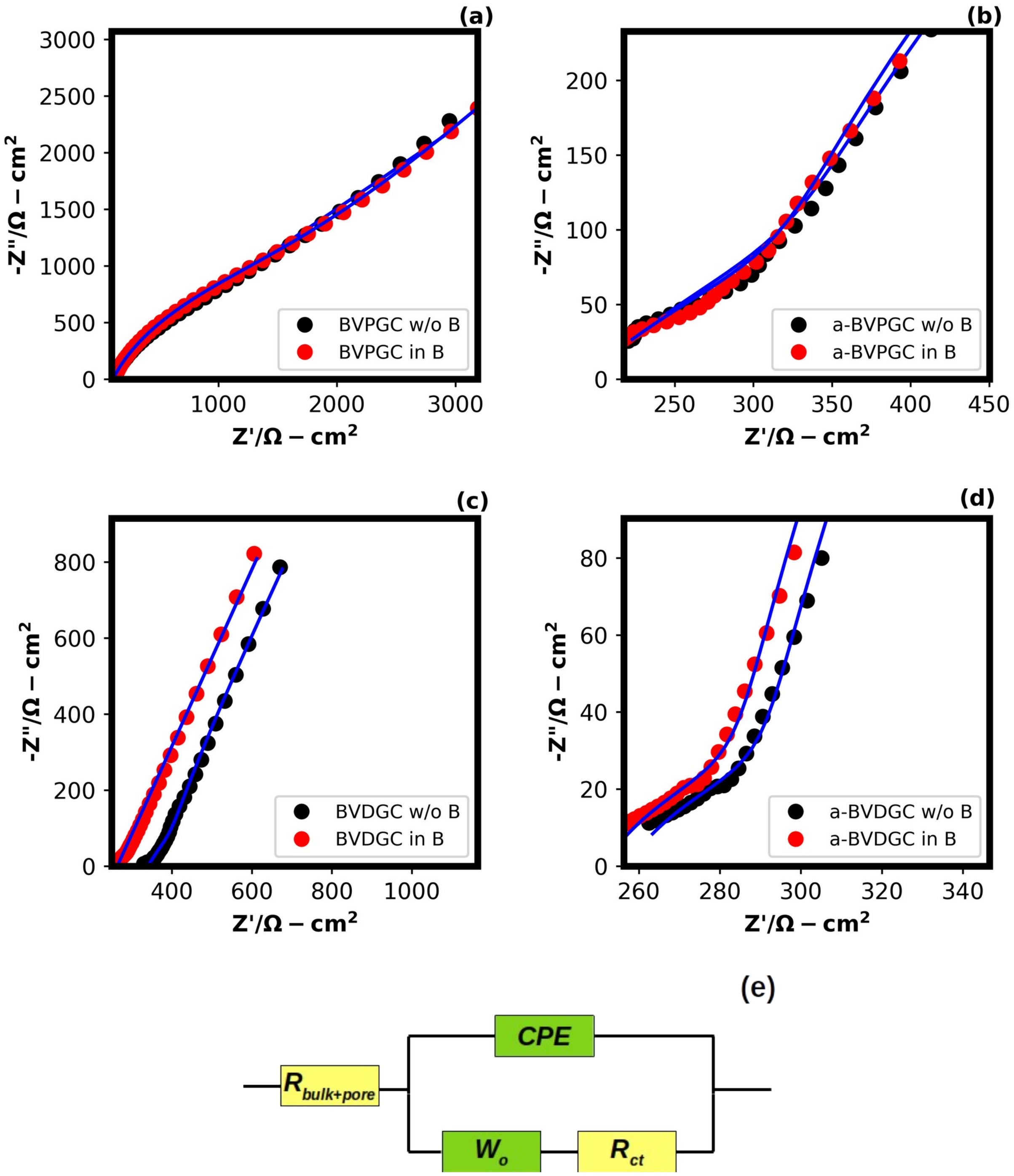

**Fig. 6**: (a-d) Fitting of equivalent circuit (blue lines) to the three electrodes cells fabricated using bougainvillea derived porous carbons. (e) Equivalent circuit used for fitting.